\documentclass[reprint, amsmath,amssymb, aps, prl,superscriptaddress,email]{revtex4-2}

\usepackage{graphicx}
\usepackage{comment}
\usepackage{amsmath}
\usepackage{mathtools}
\usepackage{amssymb}
\usepackage{amsthm}
\usepackage{bm}
\usepackage{mathtools}
\usepackage{color}
\usepackage{hyperref}
\usepackage[T1]{fontenc}
\usepackage[utf8]{inputenc}
\usepackage{braket}
\usepackage{gensymb}
\usepackage[version=4]{mhchem}

\usepackage{soul}

\begin{document}

\preprint{APS/123-QED}

\title{Molecular dynamics in Rydberg tweezer arrays: Spin-phonon entanglement and Jahn-Teller effect}

\author{Matteo Magoni}
\affiliation{Institut f\"ur Theoretische Physik, Universit\"at Tübingen, Auf der Morgenstelle 14, 72076 T\"ubingen, Germany}
\author{Radhika Joshi}
\affiliation{Institut f\"ur Theoretische Physik, Universit\"at Tübingen, Auf der Morgenstelle 14, 72076 T\"ubingen, Germany}
\author{Igor Lesanovsky}
\affiliation{Institut f\"ur Theoretische Physik, Universit\"at Tübingen, Auf der Morgenstelle 14, 72076 T\"ubingen, Germany}
\affiliation{School of Physics and Astronomy and Centre for the Mathematics and Theoretical Physics of Quantum Non-Equilibrium Systems, The University of Nottingham, Nottingham, NG7 2RD, United Kingdom}
\date{\today}

\begin{abstract} 
Atoms confined in optical tweezer arrays constitute a platform for the implementation of quantum computers and simulators. State-dependent operations are realized by exploiting electrostatic dipolar interactions that emerge, when two atoms are simultaneously excited to high-lying electronic states, so-called Rydberg states. These interactions also lead to state-dependent mechanical forces, which couple the electronic dynamics of the atoms to their vibrational motion. We explore these vibronic couplings within an artificial molecular system in which Rydberg states are excited under so-called facilitation conditions. This system, which is not necessarily self-bound, undergoes a structural transition between an equilateral triangle and an equal-weighted superposition of distorted triangular states (Jahn-Teller regime) exhibiting spin-phonon entanglement on a micrometer distance. This highlights the potential of Rydberg tweezer arrays for the study of molecular phenomena at exaggerated length scales.
\end{abstract}

\maketitle

\textbf{Introduction ---} Recent progress in controlling ultra cold atomic gases allows the preparation of atomic arrays with virtually arbitrary geometry~\cite{endres2016,barredo2016}. This technological advance is at the heart of recent breakthroughs in the domains of quantum simulation and quantum computation~\cite{Jaksch2000,Bloch2012,Yang_Wang2016,Han_2016,Su2016,Gross2017,Petrosyan2017,Shi2017,Kumar2018,Kaufman2021,Cohen2021,Jandura2022,Graham2022,Pagano2022}. Key for the latter applications is the utilization of atomic Rydberg states in which atoms interact via electrostatic dipolar interactions~\cite{Saffmann2010,Adams_2019,Beterov_2020,Wu_2021}. This mechanism underlies the experimental implementation of many-body spin Hamiltonians with variable interaction range and geometry~\cite{Singer_2005,Pohl_2009,Sandia_2016,Labuhn2016,Letscher2017,Ott_2018,Ohl_de_Mello2019,Kim2019,Steinert2023}. By building on this capability, a number of recent works have studied the dynamics of quantum correlations in many-body systems~\cite{Lienhard2018,Bluvstein2022}, critical behavior near phase transitions~\cite{Keesling2019,Samajdar2021} and novel manifestations of ergodicity breaking~\cite{Bluvstein2021}.

\begin{figure}
\centering
\includegraphics[width=\columnwidth]{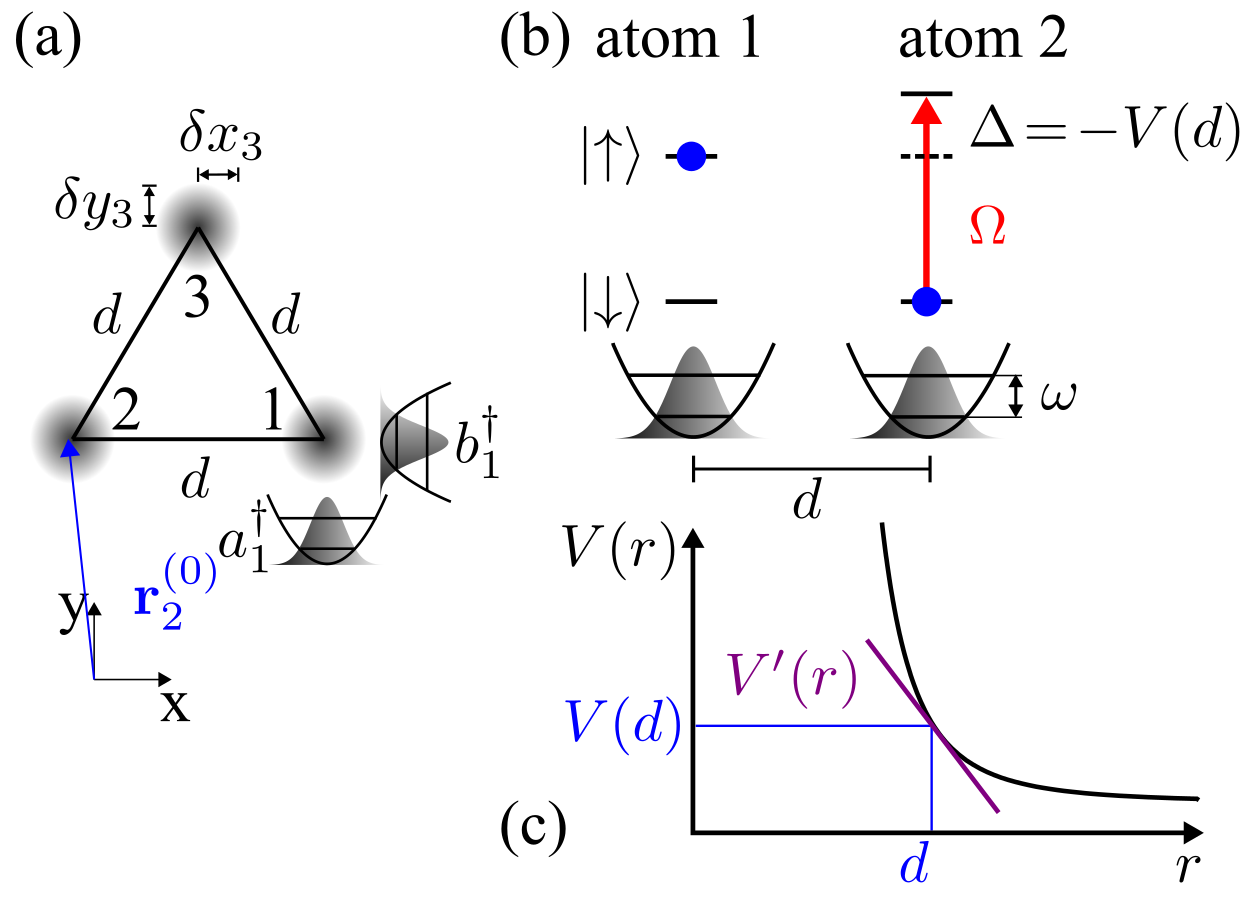}
\caption{\textbf{Artificial molecular system.}  (a) The system under consideration is formed by atoms confined in harmonic optical tweezer traps (trap frequency $\omega$) forming an equilateral triangle with side length $d$. The center of the respective traps is located at position $\mathbf{r}_j^{(0)}$. We consider motion in the $xy$-plane around the trap centers. The corresponding degrees of freedom are $\delta x_j$ ($\delta y_j$) which can be represented in terms of phonon creation operators $a_j^\dagger$ ($b_j^\dagger$). (b) Atoms are modelled as two-level systems, where $\mid\downarrow\rangle$ is the ground state and $\mid\uparrow\rangle$ the Rydberg state. Both states are coupled by a laser with Rabi frequency $\Omega$ and detuning $\Delta$ with respect to the single-atom transition frequency. When one atom is excited into a Rydberg state, the simultaneous excitation of the neighboring one requires an additional energy. This energy shift is given by the interaction potential $V$, which depends on their distance $d$. The atoms are confined in a harmonic potential with trap frequency $\omega$, which we assume to be state-independent. (c) When two atoms are simultaneously in the Rydberg state they interact with the potential $V(r)$. The gradient of the potential $V^\prime(r)$ at the interatomic distance $d$ gives rise to a force which leads to a coupling between the electronic degrees of freedom of the atoms and the vibrational dynamics within the tweezers.}
\label{fig:model}
\end{figure}
Concomitant with the strong dipolar interactions among Rydberg atoms are mechanical forces, which owed to their state-dependent nature couple the internal atomic degrees of freedom with the external motional ones~\cite{Ates2008,Wuster2010}. On the one hand, in quantum simulators and processors this mechanism causes decoherence of the electronic dynamics~\cite{Roghani2011,Keating2015,Robicheaux2021,Chew2022}; and in the extreme case they may even lead to a rapid ``explosion'' of ensembles of Rydberg atoms~\cite{Faoro2016}. On the other hand, this vibronic coupling can be exploited to implement coherent many-body interactions~\cite{Gambetta2020} and cooling protocols~\cite{Belyansky2019}, and may also enable the exploration of polaronic physics in Rydberg lattice gases dressed by phonons~\cite{Plodzien2018,Mazza2020,Magoni_phonon_dressing}. Beyond that, it enables the realization of dynamical processes that bear close resemblance to those found in molecules, but on exaggerated micrometer length scales~\cite{Shaffer2018}. The viability of this idea has been recently investigated in a theoretical work on conical intersections in an artificial molecule realized with two trapped Rydberg ions \cite{Gambetta2021}. Other examples of exotic types of molecules involving Rydberg states include ultralong-range Rydberg molecules reported in Refs.~\cite{Greene_2000,Bendkowsky2009} and Rydberg macrodimers investigated in Refs.~\cite{Boisseau_2002,Overstreet2009,Kiffner_2012,Deiglmayr2016,Hollerith2019}.

In this work we introduce and theoretically investigate an artificial molecular system which is realized in a small two-dimensional tweezer array in which atoms can be flexibly arranged. We focus on a simple setting where three trapped atoms, forming an equilateral triangle, are excited to Rydberg states under facilitation conditions. Our study, which is closely related to the physics of Rydberg aggregates~\cite{Schempp2014,Aliyu2018,Wuster2018}, establishes how the molecular spectrum is affected by vibronic couplings. It moreover reveals the emergence of a Jahn-Teller regime where the artificial molecule exhibits spin-phonon entanglement on micrometer distances. This highlights the vast possibilities offered by Rydberg arrays for studying complex dynamical processes involving coherent molecular dynamics near intersecting potential energy surfaces. Our findings also connect to recent research concerning the creation and exploitation of macroscopic quantum superposition of states in mechanical systems~\cite{Frowis2018,Weiss2021}.

\textbf{Model ---} The system we consider here is shown in Fig.~\ref{fig:model}a. The atoms form an equilateral triangle where the distance between neighboring atoms is $d$. The trapping potential within each tweezer shall be approximated by a two-dimensional (we consider the third dimension to be frozen out) isotropic harmonic trap with frequency $\omega_x = \omega_y = \omega$. Moreover, the trapping potential is assumed to be the same no matter whether an atom is in its ground state or Rydberg state. Such state-independent trapping can, for example, be achieved by operating the trapping laser at a so-called magic frequency~\cite{Zhang2011,Lampen2018,Barredo2020,Madjarov2020,Wilson2022}. Each atom is modeled as a two-level system (see Fig.~\ref{fig:model}b), where $\ket{\downarrow}$ denotes the atomic ground state and $\ket{\uparrow}$ denotes the Rydberg excited state. These two states are coupled through a laser with Rabi frequency $\Omega$ and detuning $\Delta$. Two atoms (at positions $\bm{r}_j,\bm{r}_k$) in the Rydberg state interact via a distance-dependent potential, typically of dipolar or van der Waals type $V(\bm{r}_j,\bm{r}_k)=V(|\bm{r}_j-\bm{r}_k|)$, as depicted in Fig.~\ref{fig:model}c. The Hamiltonian of the system is therefore given by ($\hbar=1$)
\begin{eqnarray}
H_\textrm{full} &=& \sum_{j = 1}^3 \left[ \Omega \hat{\sigma}^x_j + \Delta \hat{n}_j + \omega (\hat{a}^\dagger_j \hat{a}_j + \hat{b}^\dagger_j \hat{b}_j) \right]  \nonumber\\
&&+ \sum_{j = 1}^3 \sum_{k<j} V(\bm{r}_j,\bm{r}_k)\hat{n}_j\hat{n}_k,
\label{eq:ham}
\end{eqnarray}
where $\hat{\sigma}^x_j = \ket{\uparrow}_j \bra{\downarrow}_j + \ket{\downarrow}_j \bra{\uparrow}_j$ is the spin flip operator and $\hat{n}_j =  \ket{\uparrow}_j \bra{\uparrow}_j$ is the projector onto the Rydberg state of atom $j$. The operators $\hat{a}_j$ and $\hat{b}_j$ are the annihilation operators along the $x$ and $y$ directions of the two-dimensional trap holding atom $j$. The displacement of the position of atom $j$ from the center of the trap $\bm{r}_j^{(0)}$ is given by $\delta \bm{r}_j = \bm{r}_j - \bm{r}_j^{(0)} = (\delta x_j, \delta y_j)$. Assuming that the position fluctuations are small compared to the interatomic distance, $|\delta \bm{r}_j| \ll d$, we can expand the interaction potential around the equilibrium positions as
\begin{equation}
V(\bm{r}_j,\bm{r}_k) \simeq V(d) + \nabla V(\bm{r}_j,\bm{r}_k)|_{(\bm{r}_j^{(0)},\bm{r}_k^{(0)})} \cdot (\delta \bm{r}_j, \delta \bm{r}_k).
\label{eq:pot_expansion}
\end{equation}
In the following we consider the situation in which Rydberg atoms are excited under facilitation conditions~\cite{Fac1,Amthor2010,Su2017,LocFac2,Magoni_bloch_osc} as depicted in Fig. \ref{fig:model}b. Here the energy shift induced by the interaction among adjacent Rydberg atoms is cancelled by the laser detuning: $\Delta + V(d) = 0$.
\begin{figure}
\centering
\includegraphics[width=0.8\columnwidth]{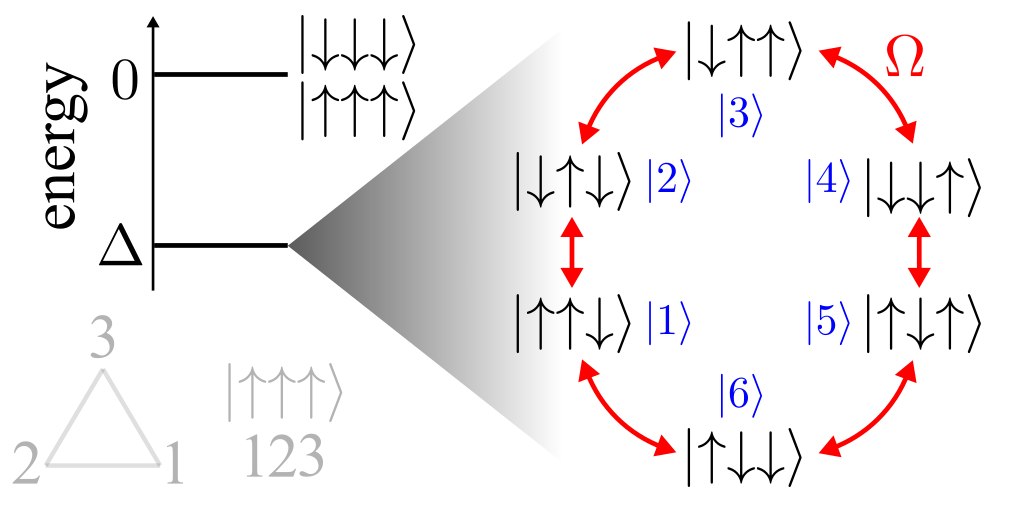}
\caption{\textbf{Tight-binding model of near-resonant states.} Under facilitation conditions, the Hilbert space of the internal dynamics of the Rydberg triangle separates into two manifolds. We are interested here in the manifold that is composed by atomic configurations of energy $\Delta$. Its basis states are coupled with Rabi frequency $\Omega$. The emerging structure corresponds to a tight-binding Hamiltonian with six sites, $|k\rangle$ with $k=1,...,6$, and periodic boundary conditions. The order of the spins appearing in the kets is indicated in the bottom left.}
\label{fig:resonant_states}
\end{figure}
Under these conditions, the Hilbert space splits into disconnected sectors that contain states with the same energy, as shown in Fig.~\ref{fig:resonant_states}. The most interesting sector is the one including the states with one or two Rydberg excitations, which have energy $\Delta$. These six near-resonant states, $|k\rangle$, belonging to this Hilbert subspace form the fictitious lattice sites of a tight-binding Hamiltonian with periodic boundary conditions, as depicted in Fig.~\ref{fig:resonant_states}. They are labeled as: $\ket{1} = \ket{\uparrow \uparrow \downarrow}$, $\ket{2} = \ket{\downarrow \uparrow \downarrow}$, $\ket{3} = \ket{\downarrow \uparrow \uparrow}$, $\ket{4} = \ket{\downarrow \downarrow \uparrow}$, $\ket{5} = \ket{\uparrow \downarrow \uparrow}$, $\ket{6} = \ket{\uparrow \downarrow \downarrow}$. To formulate the vibronic Hamiltonian on this subspace, we introduce the phonon operators $\delta x_j = x_\mathrm{ho} (\hat{a}_j + \hat{a}^\dagger_j)/\sqrt{2}$ and $\delta y_j = x_\mathrm{ho} (\hat{b}_j + \hat{b}^\dagger_j)/\sqrt{2}$, with $x_\mathrm{ho}=1/\sqrt{m\omega}$ being the harmonic oscillator length. This yields
\begin{eqnarray}
H_\mathrm{res} &=& \omega \sum_{j=1}^3 (\hat{a}^\dagger_j \hat{a}_j + \hat{b}^\dagger_j \hat{b}_j) + \Omega \sum_{k=1}^6 \left(\ket{k+1}\bra{k} + \mathrm{h.c.} \right)\nonumber\\
&&+ \kappa \sum_{j=1}^3 \left[\hat{d}_j^{a} (\hat{a}_j + \hat{a}_j^\dagger) + \hat{d}_j^{b} (\hat{b}_j + \hat{b}_j^\dagger) \right].
\label{eq:H_res}
\end{eqnarray}
\begin{figure*}
\centering
\includegraphics[width=2\columnwidth]{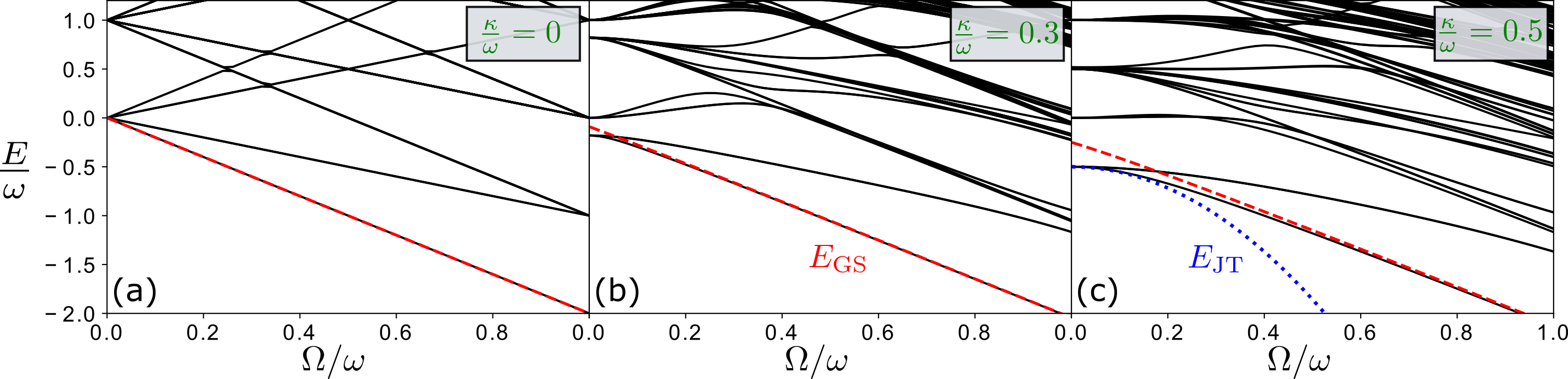}
\caption{\textbf{Energy spectrum.} Low-lying molecular energy levels as a function of the laser Rabi frequency $\Omega$ for different values of the electron-phonon coupling strength $\kappa$. The red dashed lines show the ground state energy obtained by second order perturbation theory: $E_\mathrm{GS} = E_\mathrm{GS}^{(0)}+E_\mathrm{GS}^{(2)}$. Note, that this perturbative result is valid only in the limit $\Omega\gg|\kappa|$. The blue dotted line shows the ground state energy obtained by second order perturbation theory in $\Omega$: $E_\mathrm{JT} = E_\mathrm{JT}^{(0)}+E_\mathrm{JT}^{(2)}$. The energy levels are obtained by exact diagonalization of Hamiltonian~\eqref{eq:H_res}, where we have truncated the maximum occupation number of each of the six oscillator modes at $3$.}
\label{fig:spectrum}
\end{figure*}
The first term in the first line represents the free evolution of the vibrations of the atoms in the $x$ and $y$ directions. The second term, proportional to $\Omega$, governs the tight-binding dynamics in the electronic Hilbert space (see Fig. \ref{fig:resonant_states}). The second line represents the coupling Hamiltonian between the vibrational and electronic dynamics. This coupling is parameterized by the constant 
\begin{equation}
\kappa = \left.\frac{x_\mathrm{ho}}{\sqrt{2}}\frac{\partial V(r)}{\partial r}\right|_{r=d},
\end{equation}
which is proportional to the gradient of the interaction potential at the equilibrium distance $d$. This mechanical force, which arises from the Rydberg interaction, displaces the atoms from the centers of the traps. The displacement is state-dependent which is manifest through the operators $\hat{d}_j^{a/b}$, which depend on the projectors $P_m= |m\rangle\langle m|$:
$\hat{d}_1^{a} = P_1 + \frac{1}{2} P_5$, $\hat{d}_2^{a} = - (P_1 + \frac{1}{2} P_3)$, $\hat{d}_3^{a} = \frac{1}{2}(P_3-P_5)$ as well as $\hat{d}_1^{b} = -\frac{\sqrt{3}}{2} P_5$, $\hat{d}_2^{b} = -\frac{\sqrt{3}}{2} P_3$ and $\hat{d}_3^{b} = \frac{\sqrt{3}}{2} (P_3 + P_5)$.

\textbf{Energy spectrum ---} In the following we consider the case $|\kappa| \ll \omega, \Omega$, which can be studied within a perturbative analysis. The unperturbed eigenstates are products of the Fock states $\ket{n_1^x,n_2^x,n_3^x;n_1^y,n_2^y,n_3^y}$, with occupation numbers $n_k^\alpha$ (corresponding to the number of quanta in each vibrational degree of freedom), and the eigenstates of the electronic tight-binding Hamiltonian. Specifically, the unperturbed ground state is $\ket{\mathrm{GS}}^{(0)} = \ket{\mathrm{GS}_\textrm{elec}} \ket{0,0,0;0,0,0}$, with
\begin{equation}
\ket{\mathrm{GS}_\textrm{elec}} = \frac{1}{\sqrt{6}} \sum_{m=1}^6 (-1)^m\ket{m},
\label{eq:GS_elec}
\end{equation}
and with eigenenergy $E_\mathrm{GS}^{(0)} = -2 \Omega$. Using perturbation theory up to second order in $\kappa$, the correction to the ground state energy is given by
\begin{equation}
E_\mathrm{GS}^{(2)} = - \frac{\kappa^2}{4} \left(\frac{1}{\omega} + \frac{1}{\omega + \Omega} + \frac{1}{\omega + 3\Omega} + \frac{1}{\omega + 4\Omega} \right),
\end{equation}
which well captures the level repulsion between the non-degenerate ground state and the excited states, as shown by the red dashed lines in Fig. \ref{fig:spectrum}. Fixing the coupling constant $\kappa$, the full energy spectrum displays two distinct regimes: for $\Omega \ll \omega$, the spectrum is split into groups of energy levels, which are separated by gaps of energy $\omega$. For $\Omega \gg \omega$, the spectrum is decomposed into groups in which each state possesses approximately the same eigenenergy with respect to the tight-binding Hamiltonian.

A second regime to consider is the one where $\Omega \ll \omega, |\kappa|$. Here the electronic tight-binding Hamiltonian can be treated as a perturbation. In this case, we diagonalize the unperturbed Hamiltonian by applying a unitary displacement operator
\begin{equation}
\hat{D} = \exp{\left\{-\frac{\kappa}{\omega} \sum_{j=1}^3 \left[\hat{d}_j^{a} (\hat{a}_j^\dagger - \hat{a}_j) + \hat{d}_j^{b} (\hat{b}_j^\dagger - \hat{b}_j) \right] \right\}}
\end{equation}
to Hamiltonian~\eqref{eq:H_res}, thereby obtaining
\begin{equation}
\hat{D}^\dagger H_\mathrm{res} \hat{D} = H_0 + V,
\end{equation}
where
\begin{equation}
H_0 = \omega \sum_{j=1}^3 (\hat{a}^\dagger_j \hat{a}_j + \hat{b}^\dagger_j \hat{b}_j) - 2 \frac{\kappa^2}{\omega} \left(P_1 + P_3 + P_5 \right)
\end{equation}
is diagonal in the product states $\ket{k} \ket{n_1^x,n_2^x,n_3^x;n_1^y,n_2^y,n_3^y}$, while
\begin{equation}
V = \Omega \hat{D}^\dagger \sum_{k=1}^6 \left(\ket{k+1}\bra{k} + \mathrm{h.c.} \right) \hat{D}
\label{eq:potential}
\end{equation}
is a displaced hopping operator, whose explicit expression is derived in the Supplemental Material. The unperturbed Hamiltonian $H_0$ is characterized by the presence of three degenerate ground states, each with eigenvalue $- 2 \kappa^2/\omega$, given by the phonon vacuum and the electronic states with two Rydberg excitations. This degeneracy is a manifestation of the invariance of the Hamiltonian under a 120\degree \, rotation around the center of mass of the equilateral triangle. By applying degenerate perturbation theory, one finds that the ground state degeneracy is partially lifted at second order in $\Omega$, allowing to select the right ground state from the three-dimensional ground state manifold. This is given as
\begin{eqnarray}
\ket{\mathrm{JT}}^{(0)} &=& \frac{1}{\sqrt{3}} \left(\ket{1} \Ket{-\frac{\kappa}{\omega}, \frac{\kappa}{\omega},0;0,0,0} \right. \nonumber \\
&&+ \ket{3} \Ket{0,\frac{\kappa}{2\omega},-\frac{\kappa}{2\omega};0,\frac{\sqrt{3}\kappa}{2\omega},-\frac{\sqrt{3}\kappa}{2\omega}} \nonumber \\
&&+ \left. \ket{5} \Ket{-\frac{\kappa}{2\omega},0,\frac{\kappa}{2\omega};\frac{\sqrt{3}\kappa}{2\omega},0,-\frac{\sqrt{3}\kappa}{2\omega}}\right)
\label{eq:JT_GS}
\end{eqnarray}
and it has eigenenergy $E_\mathrm{JT}^{(0)} = -2 \kappa^2/\omega$. It consists in a spin-phonon entangled state, where each of the three degenerate electronic states is coupled to a different set of motional coherent states. This state represents a neat manifestation of the Jahn-Teller effect~\cite{Jahn_Teller_1,Jahn_Teller_2}, as each motional coherent state represents a possible distortion of the triangular configuration. This state is fundamentally different from the product state $\ket{\mathrm{GS}}^{(0)}$, in which the atoms remain placed at the corners of an equilateral triangle. The energy shift of the ground state due to the Rabi coupling, computed by second order perturbation theory (see Supplemental Material), is given by
\begin{equation}
E_\mathrm{JT}^{(2)} = - 2\frac{\Omega^2}{\omega} \left[\frac{\Gamma(\eta,0,-\eta)}{(-\eta e)^{\eta}} + \frac{\Gamma \left(\eta,0,-\frac{\eta}{4} \right)}{\left(-\frac{\eta e}{4} \right)^{\eta}} \right], 
\end{equation}
where $\Gamma(a,0,x) = \int_0^x t^{a-1} e^{-t} dt$ is the incomplete Gamma function and $\eta = 2 \kappa^2/\omega^2$. It quantifies the curvature of the ground state energy for small $\Omega$ shown as the blue dotted line in Fig.~\ref{fig:spectrum}c. 

\textbf{Born-Oppenheimer treatment ---}
\begin{figure}
\centering
\includegraphics[width=\columnwidth]{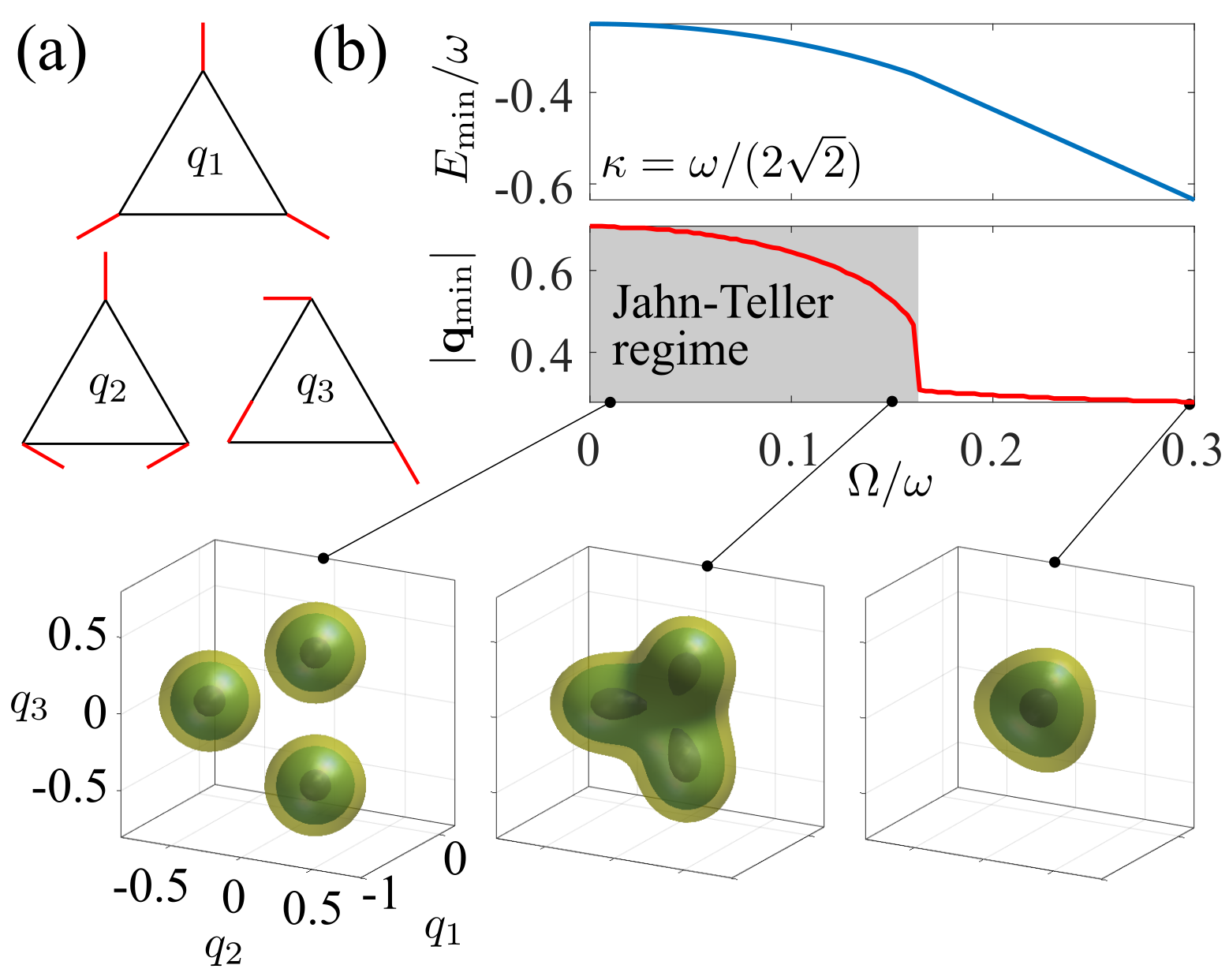}
\caption{\textbf{Born-Oppenheimer energy surfaces and structural transition.} (a) Sketch of the normal modes. The red lines indicate the distortion of the triangle associated with the $q_m$. (b) Minimum energy of the lowest Born-Oppenheimer surface $E_0(\mathbf{q})$ and position $|\mathbf{q}_\mathrm{min}|$ of the minimum. In the Jahn-Teller regime the minimum is three-fold degenerate as can be seen in the iso-energy surfaces. Lengths are given in units of the harmonic oscillator length $x_\mathrm{ho}=1/\sqrt{m\omega}$. The axes labels are the same for all the panels in the bottom.}
\label{fig:Jahn-Teller}
\end{figure}
An instructive perspective on the structural properties of the artificial triangular molecular system is obtained by analyzing its lowest energy states within the Born-Oppenheimer approximation. To this end we neglect the kinetic energy of the atoms and write the Hamiltonian in terms of the normal modes $q_m$ (see Supplemental Material), which are shown in Fig. \ref{fig:Jahn-Teller}a:
\begin{eqnarray}
H_\mathrm{BO}&=&\frac{\omega}{2} \sum_{m=1}^6 \frac{q_m^2}{x_\mathrm{ho}^2} + \sqrt{2}\kappa (P_1+P_3+P_5) \frac{q_1}{x_\mathrm{ho}} \nonumber \\&&- \frac{\kappa}{\sqrt{2}} (2 P_1 - P_3 - P_5) \frac{q_2}{x_\mathrm{ho}} - \sqrt{\frac{3}{2}} \kappa (P_3-P_5) \frac{q_3}{x_\mathrm{ho}} \nonumber \\
&& +\Omega \sum_{k=1}^6 \left(\ket{k+1}\bra{k} + \mathrm{h.c.} \right).
\end{eqnarray}
Calculation of the lowest eigenenergy of this Hamiltonian yields the ground state Born-Oppenheimer surface as a function of the normal coordinates $E_0(\mathbf{q})$. For sufficiently small values of $\Omega$ this surface has three degenerate minima, as can be seen in Fig. \ref{fig:Jahn-Teller}b. This is the Jahn-Teller regime~\cite{Kokoouline2003,Honda2006,Porras2012}, where the ground state of the full (quantum) problem is a superposition of three triangular configurations that have only one distorted side. When $\Omega$ increases the minima move towards each other until they collapse. From here onward the electronic and vibrational dynamics approximately factorise: the electronic state can by approximated by $\ket{\mathrm{GS}_\textrm{elec}}$ and the external degrees of freedom arrange in way that leads to the minimization of the projected Hamiltonian $\bra{\mathrm{GS}_\textrm{elec}}H_\mathrm{BO}\ket{\mathrm{GS}_\textrm{elec}}=\frac{\omega}{2} \sum_{m=1}^6 q_m^2/x_\mathrm{ho}^2 + \frac{\kappa}{\sqrt{2}} q_1/x_\mathrm{ho} -2 \Omega$. Here only the mode $q_1$ gets displaced, while the other two modes remain at the origin, as shown in the rightmost panel of Fig. \ref{fig:Jahn-Teller}b. Since only $q_1 \neq 0$, the displaced atoms remain at the vertices of an equilateral triangle. 

\textbf{Experimental considerations ---} Eigenstates of the artificial molecular system can be prepared from the initial state $\mid\downarrow\downarrow\downarrow\rangle$ using an adiabatic ramp, which has been already demonstrated for substantially larger Rydberg atom arrays than discussed here~\cite{Schauss2015,Ebadi2021,Scholl2021}. In the Jahn-Teller regime the ground state~\eqref{eq:JT_GS} is a superposition of three states that minimize the energy. A measurement of the Rydberg density selects one of these states, corresponding to a configuration in which the atoms form a distorted triangle. This distortion is given by $|\delta \bm{r}| = \frac{|V^\prime(d)|}{m \omega^2}$, which is equal to the classical displacement of two interacting particles confined in a harmonic potential. To estimate the distortion, we consider \ce{^{39}K} atoms held in optical tweezers with trap frequency $\omega = 2 \pi \times 70$ kHz at interatomic distance $d = 5 \, \mu$m. With the van der Waals interaction between 60$S$ Rydberg states one obtains $V^\prime(d) = 6 C_6/d^7 \simeq 6.76 \cdot 10^{-3}$ GHz $\mu$m$^{-1}$~\cite{ARC_Rydberg}, which yields a Jahn-Teller distortion of $|\delta \bm{r}| \simeq 350$ nm. The position of Rydberg atoms and the transition into the Jahn-Teller regime with decreasing $\Omega$ can thus be detected by field ionization as the created ions can then be detected with high spatial resolution ($\sim 200$ nm) as shown recently in Ref.~\cite{Zou2023}, where the vibrational dynamics of Rydberg-ion molecules was probed. An alternative way to probe the Jahn-Teller distortion is through a reconstruction of the Wigner function, as recently demonstrated in the context of trapped neutral atoms in Ref.~\cite{Winkelmann_2022} with a direct approach and in Ref.~\cite{Regal2023} with time-of-flight imaging techniques. Note, that throughout we have assumed that we operate at zero temperature, which is currently still a challenge.

\textbf{Summary and outlook ---} We studied the creation of molecular states formed in small Rydberg tweezer arrays. Their structure is dictated by the interplay between mechanical forces and coherent laser excitation, which gives rise to a crossover into a Jahn-Teller regime. Note, that throughout the paper, we have assumed that the atoms are trapped with a trap frequency that does not depend on their internal electronic state. We leave the analysis of this interesting scenario to future work. In the future it will also be interesting to investigate even more complex scenarios, such as conical intersections~\cite{Wuster2011,Hummel2021}. This would enable the experimental probing of dynamical effects, such as the impact of geometric phases or non-adiabatic couplings among Born-Oppenheimer surfaces, in a molecular system on a micrometer length scale~\cite{Gambetta2021}. Moreover, as in the Jahn-Teller regime small external fields give rise to a measurable configuration change, artificial molecular states could potentially be utilized for sensing applications.

\begin{acknowledgments}
We are grateful for financing from the Baden-W\"urttemberg Stiftung through Project No.~BWST\_ISF2019-23. We also acknowledge funding from the Deutsche Forschungsgemeinschaft (DFG, German Research Foundation) under Projects No. 428276754 and 435696605 as well as through the Research Unit FOR 5413/1, Grant No. 465199066. This project has also received funding from the European Union’s Horizon Europe research and innovation program under Grant Agreement No. 101046968 (BRISQ). We thank C. Gro{\ss} for discussions.
\end{acknowledgments}

\bibliography{bib}

\setcounter{equation}{0}
\setcounter{figure}{0}
\setcounter{table}{0}
\makeatletter
\renewcommand{\theequation}{S\arabic{equation}}
\renewcommand{\thefigure}{S\arabic{figure}}
\renewcommand{\bibnumfmt}[1]{[S#1]}
\renewcommand{\citenumfont}[1]{S#1}

\onecolumngrid
\newpage

\begin{center}
{\Large SUPPLEMENTAL MATERIAL}
\end{center}
\begin{center}
\vspace{0.8cm}
{\Large Molecular dynamics in Rydberg tweezer arrays: Spin-phonon entanglement and Jahn-Teller effect}
\end{center}

\begin{center}
Matteo Magoni$^{1}$, Radhika Joshi$^{1}$, and Igor Lesanovsky$^{1,2}$
\end{center}
\begin{center}
$^1${\em Institut f\"ur Theoretische Physik, Universit\"at Tübingen,}\\
{\em Auf der Morgenstelle 14, 72076 T\"ubingen, Germany}\\
$^2${\em School of Physics and Astronomy and Centre for the Mathematics}\\
{\em and Theoretical Physics of Quantum Non-Equilibrium Systems,}\\
{\em The University of Nottingham, Nottingham, NG7 2RD, United Kingdom}
\end{center}

We present here some details on the calculations of the main text. In the first section, we provide the derivation of the expansion of the interaction potential around the equilibrium position of the atoms and the expression of Hamiltonian~(3) of the main text. In the second section, we present the calculation of the perturbative energy correction given by Eq.~(12) of the main text. In the third section, we provide the representation of the normal modes in terms of the atom displacements that we employ in the Born-Oppenheimer treatment.

\section{Expansion of the interaction potential and derivation of Hamiltonian~(3)}
In this section we provide the explicit calculation of the expansion of the interaction potential between two atoms excited in the Rydberg state and located at positions $\bm{r}_j$ and $\bm{r}_k$. Starting from Eq.~(2) of the main text, we expand the interaction potential
\begin{equation}
V(\bm{r}_j,\bm{r}_k) \simeq V(\bm{r}_j^{(0)},\bm{r}_k^{(0)}) + \nabla V(\bm{r}_j,\bm{r}_k)|_{(\bm{r}_j^{(0)},\bm{r}_k^{(0)})} \cdot (\delta \bm{r}_j, \delta \bm{r}_k)
\label{sm:inter_pot}
\end{equation}
up to first order in the position displacement $\delta \bm{r}_j = \bm{r}_j - \bm{r}_j^{(0)} = (\delta x_j, \delta y_j)$. Since the interaction potential depends only on the distance between atoms, we rewrite it as $V(\bm{r}_j,\bm{r}_k) \equiv \widetilde{V}(r_{jk})$, where $r_{jk} = \sqrt{(x_{jk})^2 + (y_{jk})^2}$, with $x_{jk} = x_j-x_k$ and $y_{jk} = y_j-y_k$. 
Then Eq.~\eqref{sm:inter_pot} can be rewritten as
\begin{equation*}
\widetilde{V}(r_{jk}) \simeq \widetilde{V}(r_{jk}^{(0)}) + \frac{\widetilde{V}^\prime(r_{jk})}{r_{jk}} (x_{jk}, y_{jk}, -x_{jk}, -y_{jk})|_{(x_{jk}^{(0)},y_{jk}^{(0)})} \cdot (\delta x_j, \delta y_j, \delta x_k, \delta y_k).
\end{equation*}
Considering, moreover, the lattice geometry of the model, i.e. an equilateral triangle with side $d$, the three interaction potentials expanded around the equilibrium positions of the atoms read
\begin{eqnarray*}
V(\bm{r}_1,\bm{r}_2) &\simeq &\widetilde{V}(d) + \widetilde{V}^\prime(d) \delta x_{12} \nonumber \\
V(\bm{r}_2,\bm{r}_3) &\simeq &\widetilde{V}(d) + \frac{\widetilde{V}^\prime(d)}{2} \left(\delta x_{32} + \sqrt{3} \delta y_{32} \right) \nonumber \\
V(\bm{r}_3,\bm{r}_1) &\simeq &\widetilde{V}(d) + \frac{\widetilde{V}^\prime(d)}{2} \left(\delta x_{13}  - \sqrt{3} \delta y_{13}  \right),
\end{eqnarray*}
with $\delta x_{jk}= \delta x_{j}-\delta x_{k}$, etc. We now define  the phonon operators $\delta x_j = \frac{x_\mathrm{ho}}{\sqrt{2}} (\hat{a}_j + \hat{a}^\dagger_j)$ and $\delta y_j = \frac{x_\mathrm{ho}}{\sqrt{2}} (\hat{b}_j + \hat{b}^\dagger_j)$, with $x_\mathrm{ho}=1/\sqrt{m\omega}$ being the harmonic oscillator length. This yields the following expression for the expansion of the interactions
\begin{equation*}
\sum_{j = 1}^3 \sum_{k<j} V(\bm{r}_j,\bm{r}_k)\hat{n}_j\hat{n}_k = \kappa \left[\hat{d}_j^{a} (\hat{a}_j + \hat{a}_j^\dagger) + \hat{d}_j^{b} (\hat{b}_j + \hat{b}_j^\dagger) \right],
\end{equation*}
which is the second line of Eq.~(3) of the paper. Here $\kappa$ and the operators $\hat{d}_j^{a/b}$ are defined as in the main text.

\section{Expression of the displaced hopping operator~(10) and derivation of Eq.~(12)}
In this section we provide the expression of the displaced hopping operator given by Eq.~(10) of the main text and the derivation of the second order correction to the ground state energy given by Eq.~(12) of the main text. 

The displaced hopping operator (in units of $\Omega$) is given by
\begin{eqnarray}
\frac{V}{\Omega} &=& \hat{D}^\dagger \sum_{k=1}^6 \left(\ket{k+1}\bra{k} + \mathrm{h.c.} \right) \hat{D} \nonumber \\
&=& \exp{\left\{\frac{\kappa}{\omega} \sum_{j=1}^3 \left[\hat{d}_j^{a} (\hat{a}_j^\dagger - \hat{a}_j) + \hat{d}_j^{b} (\hat{b}_j^\dagger - \hat{b}_j) \right] \right\}} \left[\sum_{k=1}^6 \ket{k+1}\bra{k} + \mathrm{h.c.}  \right] \exp{\left\{-\frac{\kappa}{\omega} \sum_{j^\prime=1}^3 \left[\hat{d}_{j^\prime}^{a} (\hat{a}_{j^\prime}^\dagger - \hat{a}_{j^\prime}) + \hat{d}_{j^\prime}^{b} (\hat{b}_{j^\prime}^\dagger - \hat{b}_{j^\prime}) \right] \right\}} \nonumber \\
&=& e^{-\frac{\kappa^2}{\omega^2}} \left[\left(\ket{1}\bra{2} + \ket{1}\bra{6} \right) e^{\frac{\kappa}{\omega} \hat{a}_1^\dagger} e^{-\frac{\kappa}{\omega} \hat{a}_1} e^{-\frac{\kappa}{\omega} \hat{a}_2^\dagger} e^{\frac{\kappa}{\omega} \hat{a}_2} + \left(\ket{3}\bra{2} + \ket{3}\bra{4} \right) e^{\frac{\kappa}{2\omega} \hat{a}_3^\dagger} e^{-\frac{\kappa}{2\omega} \hat{a}_3} e^{-\frac{\kappa}{2\omega} \hat{a}_2^\dagger} e^{\frac{\kappa}{2\omega} \hat{a}_2} e^{\frac{\sqrt{3}\kappa}{2\omega} \hat{b}_3^\dagger} e^{-\frac{\sqrt{3}\kappa}{2\omega} \hat{b}_3} e^{-\frac{\sqrt{3}\kappa}{2\omega} \hat{b}_2^\dagger} e^{\frac{\sqrt{3}\kappa}{2\omega} \hat{b}_2} \right. \nonumber \\
&&+ \left. \left(\ket{5}\bra{4} + \ket{5}\bra{6} \right) e^{\frac{\kappa}{2\omega} \hat{a}_1^\dagger} e^{-\frac{\kappa}{2\omega} \hat{a}_1} e^{-\frac{\kappa}{2\omega} \hat{a}_3^\dagger} e^{\frac{\kappa}{2\omega} \hat{a}_3} e^{-\frac{\sqrt{3}\kappa}{2\omega} \hat{b}_1^\dagger} e^{\frac{\sqrt{3}\kappa}{2\omega} \hat{b}_1} e^{\frac{\sqrt{3}\kappa}{2\omega} \hat{b}_3^\dagger} e^{-\frac{\sqrt{3}\kappa}{2\omega} \hat{b}_3} + \mathrm{h.c.} \right],
\label{sm:pert}
\end{eqnarray}
where we use the definition of the operators $\hat{d}_j^{a/b}$ given in the main text and the Kermack-McCrae identity for the displacement operator $\hat{D}(\alpha) = e^{-\frac{1}{2}|\alpha|^2} e^{\alpha \hat{a}^\dagger} e^{-\alpha^* \hat{a}}$.

The unperturbed Hamiltonian given by Eq.~(9) has three degenerate ground states, namely $\ket{A} \equiv \ket{1}\ket{0,0,0;0,0,0}$, $\ket{B} \equiv \ket{3}\ket{0,0,0;0,0,0}$ and $\ket{C} \equiv \ket{5}\ket{0,0,0;0,0,0}$, with eigenvalue $E_\mathrm{JT}^{(0)} = - 2 \kappa^2/\omega$. To find perturbatively the ground state energy correction, we need to find the right superposition of these states that diagonalize the perturbation~\eqref{sm:pert}. Since all the matrix elements of the perturbation vanish, we have to diagonalize the perturbation at second order, employing second order degenerate perturbation theory. This is accomplished by diagonalizing the matrix $\mathbb{M} = -V (H_0 - E_\mathrm{JT}^{(0)})^{-1} V$ in the subspace spanned by the states $\{\ket{A},\ket{B},\ket{C}\}$. 

Let us first derive the matrix elements $m_{ij} = \braket{i|\mathbb{M}|j}$. The action of $\mathbb{M}$ on the state $\ket{A}$ is
\begin{eqnarray*}
\mathbb{M} \ket{A} &&= -V (H_0 - E_\mathrm{JT}^{(0)})^{-1} V \ket{A} \nonumber \\
&&= -V (H_0 - E_\mathrm{JT}^{(0)})^{-1} \Omega e^{-\frac{\kappa^2}{\omega^2}}(\ket{2}+\ket{6}) \sum_{n_1^x,n_2^x = 0}^\infty \frac{\left(-\kappa/\omega \right)^{n_1^x}}{\sqrt{n_1^x!}} \frac{\left(\kappa/\omega \right)^{n_2^x}}{\sqrt{n_2^x!}} \ket{n_1^x,n_2^x,0;0,0,0} \nonumber \\
&&= - V \Omega e^{-\frac{\kappa^2}{\omega^2}} (\ket{2}+\ket{6}) \sum_{n_1^x,n_2^x = 0}^\infty \frac{1}{\omega(n_1^x+n_2^x) + 2 \kappa^2/\omega} \frac{\left(-\kappa/\omega \right)^{n_1^x}}{\sqrt{n_1^x!}} \frac{\left(\kappa/\omega \right)^{n_2^x}}{\sqrt{n_2^x!}} \ket{n_1^x,n_2^x,0;0,0,0}.
\end{eqnarray*}
One can then evaluate the matrix element $m_{AA} = \braket{A|\mathbb{M}|A}$, which is given by
\begin{eqnarray}
m_{AA} &= - \Omega^2 e^{-2\frac{\kappa^2}{\omega^2}} &\left[(\bra{2} + \bra{6}) \sum_{n_1^{x \prime},n_2^{x \prime} = 0}^\infty \frac{\left(-\kappa/\omega \right)^{n_1^{x \prime}}}{\sqrt{n_1^{x \prime}!}} \frac{\left(\kappa/\omega \right)^{n_2^{x \prime}}}{\sqrt{n_2^{x \prime}!}} \bra{n_1^{x \prime},n_2^{x \prime},0;0,0,0} \right] \nonumber \\
&&\cdot \left[(\ket{2}+\ket{6}) \sum_{n_1^x,n_2^x = 0}^\infty \frac{1}{\omega(n_1^x+n_2^x) + 2 \kappa^2/\omega} \frac{\left(-\kappa/\omega \right)^{n_1^x}}{\sqrt{n_1^x!}} \frac{\left(\kappa/\omega \right)^{n_2^x}}{\sqrt{n_2^x!}} \ket{n_1^x,n_2^x,0;0,0,0} \right] \nonumber \\
&= -2 \Omega^2 e^{-2\frac{\kappa^2}{\omega^2}} & \sum_{n_1^x,n_2^x = 0}^\infty  \frac{1}{\omega(n_1^x+n_2^x) + 2 \kappa^2/\omega} \frac{\left(\kappa^2/\omega^2 \right)^{n_1^x}}{n_1^x!} \frac{\left(\kappa^2/\omega^2 \right)^{n_2^x}}{n_2^x!}.
\end{eqnarray}
Analogously, one finds all the other matrix elements as
\begin{eqnarray}
m_{BA} = m_{CA} &=& - \Omega^2 e^{-2\frac{\kappa^2}{\omega^2}} \sum_{n= 0}^\infty  \frac{1}{\omega n + 2 \kappa^2/\omega} \frac{\left[\kappa^2/(2\omega^2) \right]^{n}}{n!}; \nonumber \\
m_{BB} = m_{CC} &=& -2 \Omega^2 e^{-2\frac{\kappa^2}{\omega^2}} \sum_{n_1,n_2,n_3,n_4 = 0}^\infty  \frac{1}{\omega(n_1+n_2+n_3+n_4) + 2 \kappa^2/\omega} \frac{\left(\frac{\kappa^2}{4 \omega^2} \right)^{n_1}}{n_1!} \frac{\left(\frac{\kappa^2}{4 \omega^2} \right)^{n_2}}{n_2!} \frac{\left(\frac{3\kappa^2}{4 \omega^2} \right)^{n_3}}{n_3!} \frac{\left(\frac{3\kappa^2}{4 \omega^2} \right)^{n_4}}{n_4!}; \nonumber \\
m_{CB} &=& - \Omega^2 e^{-2\frac{\kappa^2}{\omega^2}} \sum_{n_1,n_2 = 0}^\infty  \frac{1}{\omega(n_1+n_2) + 2 \kappa^2/\omega} \frac{\left(-\frac{\kappa^2}{4 \omega^2} \right)^{n_1}}{n_1!} \frac{\left(\frac{3\kappa^2}{4 \omega^2} \right)^{n_3}}{n_3!}. 
\label{sm:sums}
\end{eqnarray}
These sums can be computed exactly. For example, the computation of the sum for $m_{BA}$ appearing in the first line of Eq.~\eqref{sm:sums} can be done by calculating the sum $\sum_{n=0}^\infty \frac{1}{n+a} \frac{b^n}{n!}$. The latter is given by
\begin{eqnarray*}
\sum_{n=0}^\infty \frac{1}{n+a} \frac{b^n}{n!} &&= \sum_{n=0}^\infty \int_0^\infty dt e^{-(n+a)t} \frac{b^n}{n!} = \int_0^\infty dt e^{-at} \sum_{n=0}^\infty \frac{(be^{-t})^n}{n!} = \int_0^\infty dt e^{-at} e^{be^{-t}} = - \int_{-b}^0 dy \frac{1}{y} e^{-a \ln\left(-b/y \right)} e^{-y} \nonumber \\
&&= (-b)^{-a} \int_0^{-b} dy y^{a-1} e^{-y} = (-b)^{-a} \Gamma(a,0,-b)
\end{eqnarray*}
where we make use of the equality $1/\gamma = \int_0^\infty dt e^{-\gamma t}$ and define the variable $y = -be^{-t}$. Despite both $(-b)^{-a}$ and $\Gamma(a,0,-b)$ being complex numbers (since $b >0$), their product, and therefore the result of the sum, is a real number. By using the same method, one computes all the matrix elements of $\mathbb{M}$ and finds that
\begin{eqnarray}
m_{AA} = m_{BB} = m_{CC} = -\frac{2\Omega^2}{\omega} (-\eta e)^{-\eta} \Gamma(\eta,0,-\eta) \nonumber \\
m_{BA} = m_{CA} = m_{CB} = -\frac{\Omega^2}{\omega} (-\eta e/4)^{-\eta} \Gamma(\eta,0,-\eta/4), 
\end{eqnarray}
where $\eta = 2 \kappa^2/\omega^2$. This shows that the diagonal entries of the matrix $\mathbb{M}$ are all equal, and that the off-diagonal entries are also all equal, but different from the diagonal entries. This is a manifestation of the symmetry of the perturbation~\eqref{sm:pert} under the permutation of the labels for the atoms. The eigenvalues of $\mathbb{M}$ provide the second order corrections to the degenerate ground states of the unperturbed Hamiltonian $H_0$. The eigenvalues are $\lambda_1 =  m_{AA} + 2 m_{BA}$ and $\lambda_2 = \lambda_3 =  m_{AA} - m_{BA}$. Since $m_{AA}, m_{BA} < 0$, the correction to the ground state energy is given by $\lambda_1$ and, using the notation of the main text, reads
\begin{equation}
E_\mathrm{JT}^{(2)} = m_{AA} + 2 m_{BA} = - 2\frac{\Omega^2}{\omega} \left[\frac{\Gamma(\eta,0,-\eta)}{(-\eta e)^{\eta}} + \frac{\Gamma \left(\eta,0,-\frac{\eta}{4} \right)}{\left(-\frac{\eta e}{4} \right)^{\eta}} \right], 
\end{equation}
which is Eq.~(12) of the main text. Moreover, one finds that the corresponding eigenstate is $\frac{1}{\sqrt{3}} \left(\ket{1}+\ket{3}+\ket{5} \right)\ket{0,0,0;0,0,0}$, which in the old undisplaced basis reads
\begin{eqnarray*}
\ket{\mathrm{JT}}^{(0)} &&= \hat{D} \left[\frac{1}{\sqrt{3}}(\ket{1}+\ket{3}+\ket{5})\ket{0,0,0;0,0,0} \right] \nonumber \\
&&= \frac{1}{\sqrt{3}} \left(\ket{1} \Ket{-\frac{\kappa}{\omega}, \frac{\kappa}{\omega},0;0,0,0} + \ket{3} \Ket{0,\frac{\kappa}{2\omega},-\frac{\kappa}{2\omega};0,\frac{\sqrt{3}\kappa}{2\omega},-\frac{\sqrt{3}\kappa}{2\omega}} \ket{5} \Ket{-\frac{\kappa}{2\omega},0,\frac{\kappa}{2\omega};\frac{\sqrt{3}\kappa}{2\omega},0,-\frac{\sqrt{3}\kappa}{2\omega}}\right),
\end{eqnarray*}
which is Eq.~(11) of the main text.

\section{Normal modes in the Born-Oppenheimer treatment}
We provide here the representation of the normal modes in terms of the displacements $\delta x_i$ and $\delta y_j$. To this end we define the vectors 
\begin{eqnarray}
\mathbf{R}&=&\left(x_1, x_2, x_3, y_1, y_2, y_3\right)\\
\mathbf{R_0}&=&\left(x^{(0)}_1,x^{(0)}_2,x^{(0)}_3,y^{(0)}_1,y^{(0)}_2,y^{(0)}_3\right)=d\left(\frac{1}{2},-\frac{1}{2},0,0,0\frac{\sqrt{3}}{2}\right)\\
\delta \mathbf{R}&=&\left(\delta x_1,\delta x_2,\delta x_3, \delta y_1, \delta y_2, \delta y_3\right),
\end{eqnarray}
which gather the atomic positions, equilibrium positions and displacements from equilibrium, respectively. These modes are obtained by diagonalizing the Hessian matrix, $\mathbb{H}$, of the interaction potential with elements
\begin{eqnarray}
\mathbb{H}_{ij} = \left.\frac{\partial^2}{\partial R_i R_j} \left(V(\mathbf{r}_1,\mathbf{r}_2)+V(\mathbf{r}_1,\mathbf{r}_3)+V(\mathbf{r}_2,\mathbf{r}_3)\right)\right|_{\mathbf{R}=\mathbf{R_0}}.
\end{eqnarray}
Its eigenstates read
\begin{eqnarray}
\mathbf{v}^{(1)}&=&\left(\frac{1}{2},-\frac{1}{2},0,-\frac{1}{2\sqrt{3}},-\frac{1}{2\sqrt{3}},\frac{1}{\sqrt{3}}\right)\\
\mathbf{v}^{(2)}&=&\left(-\frac{1}{2},\frac{1}{2},0,-\frac{1}{2\sqrt{3}},-\frac{1}{2\sqrt{3}},\frac{1}{\sqrt{3}}\right)\\
\mathbf{v}^{(3)}&=&\left(\frac{1}{2\sqrt{3}},\frac{1}{2\sqrt{3}},-\frac{1}{\sqrt{3}},-\frac{1}{2},\frac{1}{2},0\right)\\
\mathbf{v}^{(4)}&=&\left(0,0,0,\frac{1}{\sqrt{3}},\frac{1}{\sqrt{3}},\frac{1}{\sqrt{3}}\right)\\
\mathbf{v}^{(5)}&=&\left(\frac{1}{\sqrt{3}},\frac{1}{\sqrt{3}},\frac{1}{\sqrt{3}},0,0,0\right)\\
\mathbf{v}^{(6)}&=&\left(-\frac{1}{2\sqrt{3}},-\frac{1}{2\sqrt{3}},\frac{1}{\sqrt{3}},-\frac{1}{2},\frac{1}{2},0\right).
\end{eqnarray}
Defining the normal modes $q_m=\mathbf{v}^{(m)}\cdot \delta \mathbf{R}$, we can write the linear expansion of the state-dependent interaction potential as
\begin{eqnarray}
V(\mathbf{r}_1,\mathbf{r}_2)P_1+V(\mathbf{r}_1,\mathbf{r}_3)P_3+V(\mathbf{r}_2,\mathbf{r}_3)P_5&\approx& V(d)\left(P_1 + P_3 + P_5\right) \\
&& +\left. \frac{\partial V(r)}{\partial r}\right|_{r=d} \left[(P_1 + P_3 + P_5)q_1-\frac{1}{2}(2P_1 - P_3 - P_5)q_2 - \frac{\sqrt{3}}{2} (P_3 - P_5)q_3\right]\nonumber
\end{eqnarray}
The part proportional to $V(d)$ is cancelled by the detuning term due to the facilitation condition. Replacing the gradient of the potential with the definition of $\kappa$ given in the main text yields the state-dependent displacements in Eq. (13) of the paper.

\end{document}